\documentstyle[aas2pp4,psfig]{article} 
\begin{document}                  
\title  {Evaporation of accretion disks around black holes: 
the disk-corona transition and the connection to the advection-dominated accretion flow}
\author{ B.F. Liu}
\affil{Yunnan Observatory, Chinese Academy of Sciences, P.O.Box 110, Kunming 650011, China}
\author{ W. Yuan}
\affil{Yunnan Observatory, Chinese Academy of Sciences, P.O.Box 110, Kunming 650011,
China}
\affil{Max-Planck-Institut f\"ur Extraterrestrische Physik, Postfach 1603,
D-85740 Garching, Germany}
\author {F. Meyer and E. Meyer-Hofmeister}
\affil{Max-Planck-Institut f\"ur Astrophysik, Karl-Schwarzschild-Str.1,
D-85740 Garching, Germany}

\and 
 
\author { G.Z. Xie}
\affil{Yunnan Observatory, Chinese Academy of Sciences, P.O.Box 110, Kunming 650011,
China} 

\begin{abstract}
We apply the disk-corona evaporation model (Meyer \& Meyer-Hofmeister)
originally derived for dwarf novae to black hole systems. This model describes
the transition of a thin cool outer disk to a hot coronal
flow. The mass accretion rate determines
the location of this transition. For a number of well studied
black hole binaries we take the mass flow rates derived
from a fit of the advection-dominated accretion flow (ADAF) model to the observed spectra (for a review see
Narayan, Mahadevan, \& Quataert) and determine where the
transition of accretion via a cool disk to a coronal flow/ADAF would be
located for these rates.
We compare  with the observed location
of the inner disk edge, as estimated from the maximum velocity of the 
$\rm H_\alpha$ emission line. 
We find that the transition caused by evaporation
agrees with this  in stellar disks. 
We also show that the ADAF and the ``thin outer disk + corona'' 
are compatible in terms of the physics in the transition region.

\end{abstract}

\section{Introduction}
How gas falls on to a black hole has been studied intensively since
decades and various aspects have been worked out as the two-temperature model (Shapiro, Lightman, \& Eardley 1976) and the ion
torus model (Rees et al.\ 1982).
In recent years the features of an
advection-dominated accretion flow (ADAF)
(e.g.\ Narayan \& Yi 1995a, 1995b; Abramowicz et al.\ 1995, 1996; 
review Narayan, Mahadevan, \& Quataert 1998) have further been investigated. This physics has been
used successfully to model both
spectra and luminosities in black hole binaries 
and low luminosity galactic nuclei. 
Many of these investigations 
require the coexistence of an inner ADAF and an outer standard thin disk,
as proposed by Narayan et al.\ (1996).
Observations indicate that the transition from the cool thin disk
to the hot ADAF occurs at hundreds to thousands of gravitational
radii. Recently Blandford \& Begelman (1999) proposed to modify the ADAF
solutions by including a powerful wind ((adiabatic inflow-outflow 
solutions [ADIOSs]).

It is generally argued that the mass accretion rate determines the
transition radius (Esin et al.\ 1998). 
Several suggestions have been made to explain such a transition
(Narayan \& Yi 1995b, Honma 1996, Ichimaru 1977 and Igumenshchev, Abramowicz, \& Novikov  1998).
The evaporation of matter provides a natural explanation for the
transition from a cool disk to a hot coronal flow. 
With the evaporation model the location of the transition is
determined. 
This ``thin outer disk + corona'' configuration was suggested 
and investigated by Meyer \& Meyer-Hofmeister (1994). 
The evaporation model describes how
a hot self-sustained coronal layer is built up above the cool inert
disk that is fed by matter 
from the disk underneath. The vertical structure of the corona
establishes itself in the balance between heat generation, wind losses, and
radiation losses at the coronal-chromospheric transition layer.

 The aim of the present investigation is to see whether the transition
from a cool disk to a hot coronal flow can give a consistent picture
when combined with the ADAF model. The same question can be raised in
connection with ADIOSs (Blandford \& Begelman 1999) once such a 
model is worked out  in  more detail. 
In Sect.~2 we give a short description of the physics of the evaporation.
In Sect.~3 we compare our resulting transition radii with those
used for the ADAF spectral fit for a number of well studied X-ray binaries and the galactic nucleus M\,87.
We consider the physical transition to a hot flow/ADAF for disks
with relatively low accretion rates. 
Discussion and conclusion are given in Sect.~4 and Sect.~5.

\section{Implications from evaporation and the determination of the transition radius}
A cool disk in the potential well of an accreting compact object loses
mass into a corona.
This coronal evaporation flow physically results from the following
mechanism. The hot corona above the cool atmosphere conducts heat downwards
by electron conduction. At the bottom the temperature decreases from the
coronal value to a low chromospheric value, heat conduction becomes
inefficient, the thermal heat flow has to be radiated away. The
efficiency of radiation depends on the square of the particle number density.
There is an equilibrium established between heating by conduction and
cooling by radiation. In the stationary state this establishes an equilibrium
pressure at the interface between the chromosphere and the corona. 
So the coronal accretion flow continuously drains mass from the corona towards
the central object. The matter is resupplied by evaporation from the surface
of the underlying cool disk.

The evaporation rates were calculated numerically 
in detail by Liu, Meyer, \& Meyer-Hofmeister (1995) and included in
disk evolution for dwarf novae by Liu, Meyer, \& Meyer-Hofmeister (1997). 
The applicability to black hole
sources is discussed in Meyer-Hofmeister \& Meyer (1999a, 1999b). 
The model gives the evaporation
rate as function of radius and central mass. We measure mass in units
of solar masses,  $M=m {\rm M}_{\odot}$, radii in units of Schwarzschild radii,
$R=r {\rm R_s}, \  {\rm R_s}={2GM\over c^2}=2.95\times 10^5 m\, {\rm cm}$,
and mass accretion rates in Eddington units, $\dot M=\dot m\dot M_{\rm Edd},\ 
\dot M_{\rm Edd}={L_{\rm Edd}\over 0.1c^2}=1.39\times
10^{18}m\ {\rm g\,s^{-1}}$. This yields

\begin{equation}
\dot m_{\rm evap}\approx
30m^{0.17}r^{-1.17}  
\end{equation}
where $r$ denotes the inner radius of the  thin outer disk.
This rate depends on the distance to the central object. The closer
the disk reaches to the compact object, the stronger is the evaporation. 
The balance between the mass flow in the thin disk and the mass
evaporation determines where the transition into the coronal flow
occurs (see also Mineshige et al. 1998). In stationary accretion the
equality  $\dot m_{\rm evap}=\dot m$
together with  the dependence of the evaporation rate on $r$ gives the
size of the inner hole, that is the transition radius, $r_{\rm tr}$, 
\begin{equation}
r_{\rm tr}=18.3 m ^{0.17\over 1.17}\dot m^{-{1\over 1.17}} 
\end{equation}

This relation shows that the transition radius is determined by two parameters, the mass accretion rate  $\dot m$
and the mass of the central object $m$, the dependence on $m$ is quite
weak. 
In Figure\,1, we show the transition
radius as a function of the mass flow
rate, determined according to Eq.(2), for rates $\dot m$ relevant for
stellar black hole accretion.

\section{Consistency of the thin outer disk + corona model and the
ADAF model}
\subsection{Methods of comparison}
We discuss whether the two concepts, the transition from a cool outer
disk to a hot coronal flow and an ADAF close to the black hole,
 put together can provide a
consistent picture of the physics in accretion disks around black holes
and in galactic nuclei.
We assume here that all matter that evaporates to a coronal flow
(except for a small fraction of wind loss) will flow toward the inner  ADAF
since the transition radius is always smaller than the capture radius of 
the black hole.
What can be compared or proven together with the observational data?
For several sources the observed maximum velocity of the
${\rm H}_\alpha$ emission line is known and interpreted as belonging
to the inner
edge of the thin disk. For  chosen central mass and inclination the transition radius
${r_{\rm {tr}}}^{\rm obs}$ follows. Two ways of comparison are possible:

(1) If for a source the mass flow rate is known from the
ADAF spectral fit (which uses the transition radius ${r_{\rm tr}}^{\rm
obs}$)
this value ${ \dot m}^{\rm ADAF}$ can be taken to compute the
transition radius ${r_{\rm tr}}$ from the thin outer disk + corona model. This
comparison is performed in this letter for a number of
well studied  stellar and galactic X-ray sources. First
results for the soft X-ray black hole transient sources A0620--00 and
V404 Cyg were derived by Meyer (1999).

(2) If a complete evolution of the stellar disk during quiescence including
evaporation in a consistent way is computed, the mass flow rate $
\dot m^{\rm {corona}}$ and the transition radius ${r_{\rm tr}
}$ are known at all times. These values can then be
compared with ${r_{\rm tr}}^{\rm obs}$ and ${ \dot m}^{\rm ADAF}$
(Meyer-Hofmeister \& Meyer 1999a, 1999b). Such
a comparison takes into account further constraints as outburst recurrence
time of the binary and the amount of matter stored in the disk between
the outbursts, estimated from the outburst energy. Meyer-Hofmeister \&
Meyer (1999a, 1999b) followed the disk evolution for A0620-00 and
found agreement for their values ${ \dot m}$ and ${r_{\rm tr}}$
with ${r_{\rm tr}}^{\rm obs}$ and ${ \dot m}^{\rm ADAF}$.

\subsection{Comparison with observations}
In Table\,1 we list all the black hole X-ray binaries 
together with one example of galactic nuclei, M\,87, 
for which observations provide an
estimate for the transition radius and for which the accretion rates have
been derived using the ADAF model, 
and we compare these with our results for $r_{\rm tr}$. 
We plot transition radii as a function
of mass accretion rate in Figure\,1. 

A0620-00: The range of possible black hole masses of the soft X-ray transient
A0620-00 is $m$=4.4 to 12, depending on
the orbital inclination $i$, often $m=6.1$ is taken. 
 The transition radius was
estimated on the basis of the largest velocity, $v_{\rm max}$, seen in the
$H_\alpha$ emission line from the thin disk, 
$r_{\rm tr}^{\rm obs}={1\over 2}\left({c\sin i \over v_{\rm max}}\right)^2$. 
This value should be an upper limit for the transition to an ADAF.
For $m=6.1$ a satisfactory fit of the spectrum is found with
the value $\dot m=9.7\times 10^{-4}$ (Narayan et
al. 1997, model 1). For this value,  the transition
radius $\log r_{\rm_{tr}}=3.95$ follows from our model, close to the observationally
derived value $\log r_{\rm_ {tr}}^{\rm obs}=3.8$. 

V404 Cyg:
The parameters of V404 Cyg are well constrained as $m=12$, $i=56^{\circ}$. 
The $H_\alpha$ emission line indicates a maximal transition radius
$\log r_{\rm tr}^{\rm obs}=4.4$. A typical accretion rate determined by the ADAF fit is $\dot m =4.6\times 10^{-3}$
(Narayan et al.\ 1997, model 1 and  6), for which we  derive the transition radius $\log r_{\rm tr}= 3.4$.  
A change of rates $\dot m $ by a factor of about 2 gives also  a satisfactory 
spectral fit. In Narayan et
al. (1998, Fig.\,8) a smaller value is given, $\dot m =2\times
10^{-3}$, for which we get $\log r_{\rm tr}=3.7$.  

GRO J1655-40:
For this system the primary mass $m=7$ is well determined, but the
derivation of a transition radius from the $\rm H_\alpha$ emission
line profiles is difficult because of the blended spectrum of the secondary. 
During the early outburst  GRO J1655-40 showed a 6 day delay of the X-rays
compared to the optical radiation, a very interesting feature in
connection with evaporation. Such a long  delay, analogous to the UV
delay in dwarf novae, 
can not be explained by simple thin disk instability models. As
pointed out by Hameury et al.\ (1997), a thin disk which reaches inward
to only a certain transition radius can explain the delay. Their
computations include evaporation in a simplified manner and model the
X-ray delay. The transition radius is estimated as $r_{\rm
tr}=5\times 10^3$.
It is clear that the transition of the cool disk to a
coronal flow, which is also present  at the onset of the outburst,
can afford to have this hole first filled in with matter by diffusion
before the change of the hot state can proceed inwards to the X-ray
emitting region.
A detailed modeling of the quiescent state spectrum based on the
``ADAF plus a thin disk model'' (Hameury et al. 1997) yielded
$\dot m$ ranging from $3.5$ to $3.8\times 10^{-3}$.
This gives $\log r_{\rm tr}\approx 3.5$, which is close to
the value $\log r_{\rm tr}$=3.7 estimated from outer-disk-stability
arguments by Hameury et al. (1997) and is in agreement with the diffusion
time necessary to explain the observed delay of X-rays.

M\,87:
We calculated numerically, in stead of Eq.\,(2) for stellar objects,  
the transition radius of this object as a preliminary application of 
the thin outer disk + corona model to galactic nuclei. 
Reynolds et al.\ (1996) studied advection-dominated accretion in the
massive black hole M\,87. 
The fit to the broad band spectrum for\
$m=3 \times 10^9$ is satisfactory for values $\dot m = 10^{-3.5}$ to
$10^{-3}$. Again, a thin disk was not taken into account.
For $\dot m \approx 10^{-3}$ our numerical calculations show  
a transition radius $\log r_{\rm tr}$=3.9. 
Reynolds et al.\ (1996) had taken a much smaller outer radius,
$\log r_{\rm tr}$=3, pointing out that their fit of the spectrum is
rather insensitive to the outer radius of the advection-dominated region.
This is due to the lower temperatures  further out. 
But spectra obtained with the Faint Object Spectrograph on the
Hubble Space Telescope show emission lines broadened to FWHM $\approx
1700 {\rm km\,s^{-1}}$, interpreted as from a thin disk of ionized gas in
Keplerian rotation (Harms et al.\ 1994), images  of the disk show an 
inclination angle $i=42^{\circ}$ (Ford et al.\ 1994). This  indicates that
the thin disk reaches at a distance of $\log r=3.84$, which is close to 
what is expected from our estimates.

\subsection{Compatibility of the two models in the transition region}
In the transition region shown above, $r_{tr} > 10^3$,  electrons are still 
efficiently coupled with protons, and the ADAF can be treated as 
one-temperature and quasi-Keplerian rotation.
Then the vertical averaged equations governing 
the ADAF are (Narayan et al.\ 1998),
\begin{equation}
{d\over dR}\left(\rho RHv_r\right)=0
\end{equation}
\begin{equation}
c_s \approx \Omega H
\end{equation}
\begin{equation}
v_r {d\left(\Omega R^2\right)\over dR}={1\over \rho R H}{d\over dR}\left(\nu \rho R^3H{d\Omega\over dR}\right)
\end{equation}
\begin{equation}
\rho v_r T{ds\over dR}=q^+-q^-
\end{equation}
where all the quantities have their standard meanings.

On the other hand, at the transition radius there is no cool disk 
and the evaporation doesn't work any more. 
Thus, the vertical component of gas velocity and the heat conduction 
approach zero, the equations in the thin outer disk + corona model (Meyer \& Meyer-Hofmeister 1994) are then 
simplified as,
\begin{equation}
{\bf \nabla}\cdot \rho {\bf v}=0
\end{equation}
\begin{equation}
{dP\over dz}=-\rho {GMz\over \left(R^2+z^2\right)^{3/2}} \Rightarrow c_s\approx \Omega H
\end{equation}
\begin{equation}
v_r\approx -\alpha c_s^2/\Omega R
\end{equation}
\begin{equation}{3\over 2}\alpha P\Omega-n_e n_i\Lambda \left(T\right)+{2 \rho v_r\over R}\left({v^2\over 2}+{\gamma\over \gamma-1}{P\over \rho}-{GM\over R}\right)=0
\end{equation}

Comparing these two sets of equations, one can see that 
the first two equations are the same.
Joining eq.(3) and eq.(5) and taking $\nu ={2\over 3} \alpha c_s H$, 
we obtain $v_r=-\alpha c_s^2/\Omega R$, which is
exactly the third equation of the thin outer disk + corona model. 
$q^+$ and $q^-$ in the energy equation of the ADAF correspond to the 
viscous heating rate ${3\over 2} \alpha P \Omega$
and the optically thin cooling rate 
$n_e n_i \Lambda (T)$, respectively, in the energy equation of 
the thin outer  disk + corona. 
The entropy term $\rho v_r T{ds\over dR}$ 
can be developed as follows: 
\begin{eqnarray*}
\rho v_r T{ds\over dR}&=&\rho v_r \left[{d\left({\gamma\over \gamma-1}{P\over \rho}\right)\over dR}-{1\over \rho}{dP\over dR}\right]\\
&=&\rho v_r {d\over dR} \left({v^2\over 2}+{\gamma\over \gamma-1}{P\over \rho}-{GM\over 2R}\right)
\end{eqnarray*}
where we replaced ${1\over \rho}{dP\over dR}$ by $-v_r{dv_r \over dR}$ 
from the radial Euler equation.
 In a similar way to the thin outer disk + corona, let ${d\over dR}\sim -{2\over R}$, then the energy equation of the ADAF has the same form as that of the thin outer disk + corona model except for a factor of 1/2 in the term 
${GM\over R}$.
Therefore, we can expect a smooth transition from an thin outer  disk + corona to an ADAF.

\section{Discussion}

{\it X-ray binaries: }
The dependence of the transition from corona to thin disk  on $\dot m$ allows to understand the different regimes of a cool outer disk together with a coronal flow/ADAF.
For low $\dot m$ the transition occurs at a large radius, the thin
disk is easily  evaporated and the coronal flow/ADAF is dominant in a
large region of the accretion disk (Our examples, given in the last section,
belong to this regime). For intermediate $\dot m$, the mass supply by
accretion can balance the evaporation farther in. 
At high $\dot m$, the mass supply rate may be larger than
the evaporation rate everywhere and the thin disk or a slim disk
(Abramowicz et al.\ 1988) 
then extends all the way down to the last stable orbit.
This agrees with the picture proposed by Esin et al.\ (1997, 1998) and Narayan et al.\ (1998).
Table\,2 illustrates the regimes of different
spectral states of X-ray binaries according to the relation (Eq.2) between
mass accretion rate and transition radius. There is only one
adjustable parameter, $\dot m$, that determines the
spectral state in our scenario.

{\it Galactic nuclei: }
Observations of galactic nuclei show a wide range of luminosities,
from $L <10^{37} {\rm ergs\,s^{-1}}$ for our Galactic Center to 
$10^{48} {\rm ergs\,s^{-1}}$ for luminous quasars. The 
underluminous galactic nuclei can be understood as presently accreting
via an ADAF (Narayan et al.\ 1998). Since the physics of the
evaporation process is very similar in stellar
sources and galactic nuclei, we  investigated the
applicability of the thin outer disk + corona model 
for the latter. 
Numerical results show that the predicted transition radius 
for M87 is consistent with observations  using  
the accretion rate derived from the
ADAF fits. We expect that the evaluation of the
transition radius for low luminosity AGNs could be possible. 

\section{Conclusions}

We investigated whether the ADAF model and the thin outer disk +
corona model 
could be a consistent description of the accretion process. It is not
clear at all, since these two concepts describe different physics in
different regions. The ADAF model allows to derive mass accretion rates
from a fit to the observed spectrum. The position of the inner edge is
estimated from the observed maximal velocity of the ${\rm H}_\alpha$ emission
lines. We take these derived mass accretion rates and evaluate the
transition between outer thin  disk and coronal flow according to disk
evaporation (Meyer \& Meyer-Hofmeister 1994). 
For both X-ray binaries and the galactic nucleus M\,87 
the deduced transition radii are in agreement with those inferred from 
observations. 
Moreover, we showed that the ADAF and the thin outer disk + corona 
are compatible in terms of the physics in the transition region.
These support both, the ADAF interpretation and the
evaporation model. A detailed, more constrained investigation for the
evolution of the cool disk in A0620-00 by Meyer-Hofmeister \& Meyer
(1999a, 1999b) confirms this agreement.

{\footnotesize
\begin{table}
\caption[]{The transition radii for  black hole-accreting systems}
\begin{flushleft}
\begin{tabular}{lllll}
\hline\hline
\noalign{\smallskip}      
Object   & $m$ & $\dot m$ & log $r_{\rm tr}$ & log
$r_{\rm tr}^
{\rm obs}$\\
\noalign{\smallskip}\hline\noalign{\smallskip}
A0620-00 & 6.1     & $9.7\times 10^{-4}$     &3.95       & $3.8$  
\\
V404 Cyg & 12      & $4.6\times 10^{-3}$     & 3.42        &  $\leq 4.4$ \\
GRO J1655-40  & 7  & $3.5\times 10^{-3}$ & 3.48        &3.7    
\\
 M 87    & $3\times 10^9$  & $10^{-3}$  & 3.90        &  $3.84$     
    
\\
\noalign{\smallskip}\hline\hline\noalign{\smallskip} 
\end{tabular}
\\
Notes to the table:\\
$m$: Black hole mass in units of solar mass\\ 
$\dot m$: Mass accretion rate in units of Eddington rate, which generally varies by a factor of 2 for the ADAF fit\\
$r_{\rm tr}$: The transition radius predicted by our model \\
$r_{\rm tr}^{\rm obs}$: The ADAF-thin disk transition radius derived from
observations\\ 
\end{flushleft}
\end{table}
}

{\footnotesize
\begin{table}
\caption[]{The relation between accretion rate and transition radius 
for various spectral states of X-ray binaries}
\begin{flushleft}
\begin{tabular}{llll}
\hline\hline
\noalign{\smallskip}      
Spectral state   & $m$ & $\dot m$ & log $r_{\rm tr}$\\
\noalign{\smallskip}\hline\noalign{\smallskip}
quiescent state &   & $\la 0.01$ & $>3$               \\
low state       & 3--12  & $0.01-0.1$   & $2-3$             \\
intermediate state& &  $\sim 0.1$        & $\sim 2$                  \\
high state      &   &  $>0.1$       & $<2$                  \\
\noalign{\smallskip}\hline\hline\noalign{\smallskip} 
\end{tabular}
\end{flushleft}
\end{table}
}

\onecolumn
\begin{figure}
\psfig{figure=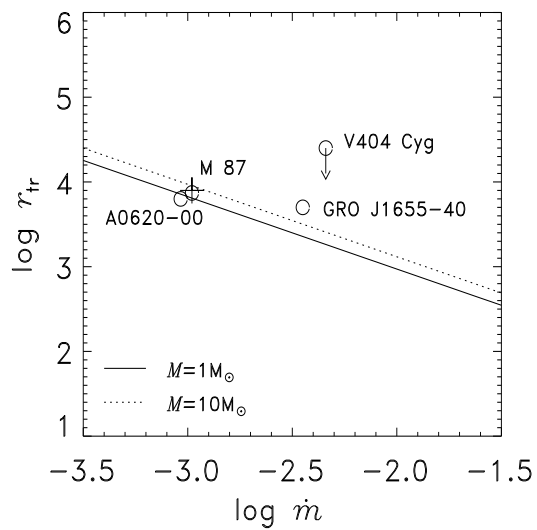,height=8.0cm,width=8.5cm,angle=0}
 \caption[]
{The lines show the transition radius $r_{\rm tr}$ as a
 function of accretion rate $\dot m$ for 
 stellar black hole masses $M$=1\,${\rm M}_{\odot}$ and 
10\,${\rm M}_{\odot}$, respectively,
according to the thin outer disk + corona model.
The circles show the transition radii derived from observations. 
The cross is the predicted transition radius 
for the galactic nucleus M\,87. 
The values are given in Table\,1.}
\end{figure}

\end{document}